\documentclass[12pt,notoc]{JHEP3} % 10pt is ignored!

\usepackage{epsfig,multicol,bbm,amsmath}

\newcommand\fverb{\setbox\fverbbox=\hbox\bgroup\verb}
\newcommand\fverbdo{\egroup\medskip\noindent%
			\fbox{\unhbox\fverbbox}\ }
\newcommand\fverbit{\egroup\item[\fbox{\unhbox\fverbbox}]}
\newbox\fverbbox

\newcommand{\nablaslash}{\not{\hbox{\kern-3pt $\nabla$}}}

\title{On relating multiple M2 and D2-branes}

\author{U.~Gran\footnote{ulf.gran@chalmers.se}, B.E.W~Nilsson\thanks{tfebn@chalmers.se} and C.~Petersson\thanks{chrpet@chalmers.se}\\
	Fundamental Physics\\
Chalmers University of Technology\\
SE-412 96 G\"oteborg, Sweden\\}

\abstract{Due to the difficulties of finding superconformal Lagrangian theories for multiple M2-branes, we will in this paper instead focus on the field equations. By relaxing the requirement of a Lagrangian formulation we can explore the possibility of having structure constants $f^{ABC}{}_D$ satisfying the fundamental identity but which are not totally antisymmetric. We exemplify this discussion by making use of an explicit choice of a non-antisymmetric $f^{ABC}{}_D$ constructed from the Lie algebra structure constants $f^{ab}{}_c$ of an arbitrary gauge group. Although this choice  of $f^{ABC}{}_D$ does not admit an obvious Lagrangian description, it does reproduce the correct SYM theory for a stack of $N$ D2-branes to leading order in $g_{YM}^{-1}$ upon reduction and, moreover, it sheds new light on the centre of mass coordinates for multiple M2-branes.}


\begin{document}

%\maketitle  IS IGNORED %%%%%%%%%%%

\section{Introduction}

Finding a model for the dynamics of multiple M2-branes is a central
problem in the quest for a better understanding of M-theory.
Building on previous work
\cite{Schwarz:2004yj,Basu:2004ed,Bagger:2006sk}, a new class of
maximally supersymmetric field equations, proposed to
describe multiple M2-branes at the conformal fixed point, was recently constructed by demanding
closure of the supersymmetry transformations
\cite{Gustavsson:2007vu,Bagger:2007jr,Bagger:2007vi}. These field
equations are parameterised by a four-index tensor $f^{ABC}{}_D$,
antisymmetric in the first three indices. One finds that in order
for the supersymmetry transformations to close these structure constants have to satisfy the so called
\emph{fundamental identity}
\begin{equation}
f^{ABC}{}_G f^{EFG}{}_D = 3f^{EF[A}{}_G f^{BC]G}{}_D  \,. \label{FI}
\end{equation}
When constructing the Lagrangian one also needs to
introduce a metric, $h^{AB}$, in order to be able to write down a
scalar with respect to the indices appearing on $f^{ABC}{}_D$. In
addition, one is forced to assume that $f^{ABCD}=f^{[ABCD]}$,
i.e.~that the four-index tensor is antisymmetric
\cite{Bagger:2007jr}. Despite the growing attention that the
Bagger-Lambert-Gustavsson theory is receiving
\cite{Gustavsson:2008dy}-\cite{Distler:2008mk},
it has proved very difficult to find solutions to the fundamental
identity using an antisymmetric $f^{ABCD}$, and so far only one
solution is known, namely $f^{ABCD}=\epsilon^{ABCD}$
\cite{Bagger:2007jr}, usually referred to as the $A_4$ theory. Note that the existence of non-trivial antisymmetric solutions $f^{ABCD}$, with index range $5,6,7$ and $8$ and positive definite metric, were ruled out in \cite{FigueroaO'Farrill:2002xg}. There is by now evidence mounting for the interpretation that the $A_4$  theory describes two coincident M2-branes \cite{Mukhi:2008ux,Lambert:2008et,Distler:2008mk}

In a recent paper by Mukhi and Papageorgakis \cite{Mukhi:2008ux}, a
mechanism for relating the proposed multiple membrane theory to a
maximally supersymmetric Yang-Mills (SYM) theory living on D2-branes was presented. Previous attempts in this direction
was made in \cite{Gustavsson:2007vu,Gustavsson:2008dy}. The
mechanism consists of making a particular choice for $f^{ABC}{}_D$,
and then in an elegant way compactify and Higgs the theory by giving a vev to
one of the scalars leading to the promotion of a non-dynamical gauge
field to a dynamical one. To obtain the SYM theory with gauge group
$SU(2)$ these authors used the $f^{ABCD}=\epsilon^{ABCD}$ mentioned
above. For gauge group $SU(N)$, i.e.~for a stack of $N>2$ branes, it was
assumed in \cite{Mukhi:2008ux} that their choice of an antisymmetric
$f^{ABCD}$, which does not solve the fundamental identity \emph{per
se}, could be completed with terms not relevant to leading order in
$g_{YM}^{-1}$ in such a way that the resulting $f^{ABCD}$ does solve
the fundamental identity. However, one can check that the $f^{ABCD}$
used to get a SYM theory with gauge group $SU(3)$ \emph{can not} be
completed in this fashion. It is therefore not clear how the procedure will work in detail for $SU(N)$, with $N>2$. However, as will be described below, by using a
set of non-antisymmetric structure constants, we are able to apply their proposed Higgs
mechanism at the level of the field equations and find the infrared limit of the SYM theory for a stack of $N$ D2-branes to leading order in $g_{YM}^{-1}$.

One motivation for studying the relation between D2
and M2-branes only at the level of the field equations, and (at least for the moment) ignore the
problem of not being able to find a Lagrangian formulation, is the great difficulty, mentioned above, of finding Lagrangians
for the supersymmetric field equations supposedly describing
multiple membranes. Furthermore, there is no \emph{a priori} reason why the infrared conformal fixed points of nonconformal $\mathcal{N}=8$ SYM theory should allow for a Lagrangian formulation. In fact, supersymmetric theories without a Lagrangian has previously been considered, e.g.~in the context of gauged supergravities \cite{Bergshoeff:2003ri}.

In this context, we note that the constraints on $f^{ABC}{}_D$
coming from demanding that the supersymmetry transformations close
are most conveniently given not in the form (\ref{FI}) of the fundamental identity, but instead as the condition
\begin{equation}
f^{[ABC}{}_G f^{E]FG}{}_D = 0 \,, \label{WFI}
\end{equation}
which, however, can be shown to be equivalent\footnote{In a previous version of this paper we assumed (\ref{FI}) and (\ref{WFI}) to be inequivalent. In section \ref{susyfe} we show that this is not the case. This fact is however of no consequence for the main conclusions of this paper.} to (\ref{FI}) as further discussed in section \ref{susyfe}. Throughout the paper we will be
careful in stating what extra assumptions we make in addition to
requiring closure of the supersymmetry algebra as some of these
assumptions might be possible to relax.

We present an explicit non-antisymmetric choice for $f^{ABC}{}_D$
in order to exemplify the above discussion. This gives rise to field
equations that, after using to the Higgs mechanism
in \cite{Mukhi:2008ux}, correspond to the SYM theory for a stack of $N$ D2-brane with an
arbitrary (compact and semi-simple) gauge group. Thus, although this choice of $f^{ABC}{}_D$ inserted into the field equations give equations which do not admit an obvious Lagrangian description, it does reproduce the infrared limit of the nonconformal D2-brane SYM theory.

Furthermore, in
contrast to previous attempts, the $U(1)$ degrees of freedom we find
on the D2-branes satisfy free field equations to all orders in
$g_{YM}^{-1}$ and thus constitute a natural candidate for the centre of mass
multiplet for multiple M2-branes. It is interesting to note that in
our approach the centre of mass coordinates arise already
at the level of the M2-branes\footnote{This has also been
noticed by Gustavsson in \cite{Gustavsson:2008dy}.}, thereby shedding
new light on the problem with the M-theory translation invariance discussed in
\cite{Mukhi:2008ux}. However, although the centre of mass modes obey free dynamics, they
do appear in the other field equations, thereby explaining why a Lagrangian, if
it exists, is not easily constructed.

The paper is organised as follows. In section \ref{susyfe} we
reanalyse the conditions that the structure constants must satisfy
in order to obtain closure of the supersymmetry transformations. We
then discuss the extra conditions needed for a straightforward
construction of a Lagrangian. A set of non-antisymmetric
structure constants relating D2 and M2-branes for any number of
branes at the level of the field equations is given in section
\ref{D2M2}. Finally, section \ref{concl} contains some closing
comments.

\section{Supersymmetric field equations}\label{susyfe}

A new class of supersymmetric field equations was recently constructed in \cite{Gustavsson:2007vu} and \cite{Bagger:2007jr}, using different but equivalent formulations \cite{Bagger:2007vi}, by demanding closure of the supersymmetry transformations, which we briefly review below. In this paper we will use the notation and conventions of \cite{Bagger:2007jr}.

Consider the fields $X_A^I,\Psi_A$ and $\tilde A_\mu{}^B{}_A$, where $X_A^I$ is an $SO(8)$ vector, $\Psi_A$ a chiral spinor and $\tilde A_\mu{}^B{}_A$ a non-dynamical gauge potential transforming as a vector under $SO(2,1)$. Indices $A,B,\ldots$ refer to an unspecified algebra defined by the structure constants $f^{ABC}{}_D$. The on-shell supersymmetry transformations are
\begin{eqnarray}
\delta X_A^I &=& i \bar \epsilon \Gamma^I \Psi_A \,,
\nonumber\\
\delta \Psi_A &=& D_\mu X_A^I \Gamma^\mu \Gamma_I \epsilon - \frac{1}{6} X_B^I X_C^J X_D^K f^{BCD}{}_A \Gamma_{IJK}\epsilon \,,
\\
\delta \tilde A_\mu{}^B{}_A &=& i \bar \epsilon \Gamma_\mu \Gamma_I X_C^I \Psi_D f^{CDB}{}_A \nonumber \,.
\end{eqnarray}
Demanding that these supersymmetry transformations close into a translation and a gauge transformation, the following field equations are obtained
\begin{eqnarray}
0 &=& \Gamma^\mu D_\mu \Psi_A + \frac{1}{2} \Gamma_{IJ} X_C^I X_D^J \Psi_B f^{CDB}{}_A \,,
\nonumber\\
0 &=& D^2 X_A^I -\frac{i}{2}\bar\Psi_C \Gamma^I{}_J X_D^J \Psi_B f^{CDB}{}_A + \frac{1}{2} f^{BCD}{}_A f^{EFG}{}_D X_B^J X_C^K X_E^I X_F^J X_G^K \,, \label{FE}
\\
0 &=& \tilde F_{\mu\nu}{}^B{}_A + \varepsilon_{\mu\nu\lambda} (X_C^J D^\lambda X_D^J + \frac{i}{2} \bar \Psi_C \Gamma^\lambda \Psi_D) f^{CDB}{}_A \,, \nonumber
\end{eqnarray}
where the covariant derivative and field strength are defined as
\begin{eqnarray}
(D_\mu X)_A &=& \partial_\mu X_A -\tilde A_\mu{}^B{}_A X_B \,,
\nonumber\\
\tilde F_{\mu\nu}{}^B{}_A &=& -2 \left(\partial_{[\mu} \tilde A_{\nu]}{}^B{}_A + \tilde A_{[\mu}{}^B{}_{|C|} \tilde A_{\nu]}{}^C{}_A \right) \,. \label{tFdet}
\end{eqnarray}
For gauge invariance one also needs to impose the relation
\begin{equation}
\tilde A_\mu{}^B{}_A = A_{\mu CD} f^{CDB}{}_A \,, \label{Aconstr}
\end{equation}
which implies a truncation of the degrees of freedom in $\tilde A_\mu{}^B{}_A$. That this is consistent with the field equations follows from the fact that using the fundamental identity one can define
\begin{equation}
F_{\mu\nu CD} := -2 \left( \partial_{[\mu} A_{\nu] CD} - A_{[\mu|EF} f^{EFG}{}_{[C} A_{|\nu]|G|D]}    \right) \,, \label{Fwot}
\end{equation}
consistent with (\ref{tFdet}), satisfying $\tilde F_{\mu\nu}{}^B{}_A = F_{\mu\nu CD} f^{CDB}{}_A $. In addition, the Bianchi identity for the gauge field is satisfied
\begin{equation}
\varepsilon^{\mu\nu\lambda} D_\mu \tilde F_{\nu\lambda}{}^B{}_A = 0 \,.
\end{equation}

When deriving the above field equations the requirement that $[\delta_1,\delta_2]\tilde A_\mu{}^B{}_A$ closes on-shell implies that $f^{ABC}{}_D$ satisfy (\ref{WFI}), or alternatively (\ref{FI}). That these conditions are indeed equivalent can be seen as follows.
By writing out (\ref{WFI}) as 
\begin{equation}
f^{ABC}{}_G f^{EFG}{}_D - 3 f^{E[AB}{}_G f^{C]FG}{}_D = 0
\end{equation}
and applying the condition (\ref{WFI}) (antisymmetrised in $E,A,B,F$) to the second term, one obtains (\ref{FI}).

The possibility of integrating the equations of motion to a Lagrangian clearly requires the existence of a metric $h^{AB}$ in order to form scalars. In addition, the only way to obtain a Lagrangian known so far is to also require $f^{ABCD}$ to be totally antisymmetric. In that case, the equations of motion above are obtained from the following Lagrangian \cite{Bagger:2007jr}
\begin{eqnarray}
{\cal L} &=& -\frac{1}{2}(D_\mu X^{AI})(D^\mu X_A^I) + \frac{i}{2} \bar \Psi^A \Gamma^\mu D_\mu \Psi_A + \frac{i}{4} \bar\Psi_B \Gamma_{IJ} X_C^I X_D^J \Psi_A f^{ABCD}
\\
&&-V +\frac{1}{2}\varepsilon^{\mu\nu\lambda}\left( f^{ABCD}A_{\mu AB}\partial_\nu A_{\lambda CD} + \frac{2}{3} f^{CDA}{}_G f^{EFGB} A_{\mu AB}  A_{\nu CD} A_{\lambda EF}  \right) \,,\nonumber
\end{eqnarray}
where
\begin{equation}
V = \frac{1}{12} f^{ABCD}f^{EFG}{}_D X_A^I X_B^J X_C^K X_E^I X_F^J X_G^K \,.
\end{equation}
After varying this Lagrangian the free gauge index must always sit on the last position of $f^{ABCD}$ in order to match with the field equations. Assuming $f^{ABCD}$ to be antisymmetric takes care of this problem.

We will now consider a particular example where there is no obvious Lagrangian formulation based on the following structure constants, where we have split the gauge indices according to $A=\{a,\phi\}$,
\begin{equation}
f^{\phi a b}{}_c = f^{ab}{}_c \,, \qquad f^{\phi a b}{}_\phi = f^{a b c}{}_\phi = f^{a b c}{}_d = 0 \,, \label{ansats}
\end{equation}
where $f^{ab}{}_c$ are the structure constants of a (compact and semi-simple) Lie algebra, see appendix \ref{useful} for details. Note in particular that this $f^{ABC}{}_D$ in \emph{non-antisymmetric} (without a metric to raise the $D$ index there is not even a notion of total antisymmetry). This form of $f^{ABC}{}_D$ has also been considered in \cite{Gustavsson:2008dy,Awata:1999dz}. In the next section we will discuss the implications of this $f^{ABC}{}_D$ at the level of equations of motion.

\section{Relating multiple M2 and D2-branes}\label{D2M2}

We will in this section investigate the physics of the explicit $f^{ABC}{}_D$ introduced in (\ref{ansats}) and, in particular, study the possible relation to D2-brane physics using the Higgs mechanism introduced recently by Mukhi and Papageorgakis \cite{Mukhi:2008ux}. In addition to focusing on the field equations, the main difference of our analysis compared to that of \cite{Mukhi:2008ux} is that we choose $f^{ABC}{}_D$ as given in (\ref{ansats}) which, even for the $SU(2)$ case, is different from the one used in \cite{Mukhi:2008ux}.

The choice of $f^{ABC}{}_D$ in (\ref{ansats}) has several special features. The condition $f^{ABC}{}_\phi = 0$ implies that the interaction terms for $X_\phi^I$ and $\Psi_\phi$ in (\ref{FE}) vanish. Also, (\ref{Aconstr}) implies that $\tilde A_\mu{}^A{}_\phi=0$ so that the covariant derivatives acting on $X_\phi^I$ and $\Psi_\phi$ reduce to ordinary derivatives. Then $X_\phi^I$ and $\Psi_\phi$ obey free dynamics and these fields will, as explained below, give rise to the $U(1)$ centre of mass degrees of freedom in the SYM theory. But, as these fields obey free dynamics already before we apply the Higgs mechanism and break the $SO(8)$ covariance, they could also be interpreted as the centre of mass degrees of freedom for multiple M2-branes. Note, however, that $X_\phi^I$ and $\Psi_\phi$ will appear in the equations of motion for the other fields in (\ref{FE}), thereby obstructing the possibility of obtaining a Lagrangian in a straightforward manner.

Let us now perform the Higgsing procedure along the lines of \cite{Mukhi:2008ux}, by giving a vev to the scalar $X_\phi^8$,
\begin{equation}
\label{fluct}
X_\phi^8 =<X_\phi^8> + x_\phi^8 = g_{YM}+ x_\phi^8~.
\end{equation}
From the last equation in (\ref{FE}) with $\tilde F_{\mu\nu}{}^b{}_a$,  we can algebraically solve for $\tilde A_\mu{}^\phi{}_a$:
\begin{equation}
\tilde A_\mu{}^\phi{}_a = \frac{1}{X_\phi^I X_\phi^I} \left( \frac{1}{2\lambda} \varepsilon_{\mu}{}^{\nu\rho} \tilde F_{\nu\rho}{}^c{}_b f_a{}^b{}_c + X_\phi^J \nabla_\mu X_a^J - X_a^J \partial_\mu X_\phi^J + i \bar \Psi_\phi \Gamma_\mu \Psi_a  \right)
\end{equation}
where $\lambda$ is defined in (\ref{lambda}). By substituting this expression back into the field equations, rescale the fields $X$ and $\Psi$ according to their canonical dimension in the world volume theory of the D2-branes ($X,\Psi) \to (X/g_{YM},\Psi/g_{YM}$) and keeping terms to leading order in $g_{YM}^{-1}$, we get
\begin{eqnarray}
\label{D2eqs}
0 &=& \Gamma^\mu \partial_\mu \Psi_\phi \,,
\nonumber\\
0 &=& \Gamma^\mu \nabla_\mu \Psi_a + \Gamma_i X_b^i \Psi_c f^{bc}{}_a  + {\cal O}\left(\frac{1}{g_{YM}^2}\right)\,,
\nonumber\\
0 &=& \partial^2 X_\phi^I \,,
\\
0 &=& \nabla^2 X_a^i - \frac{i}{2} \bar \Psi_b \Gamma^i \Psi_c f^{bc}{}_a
-  f^{cd}{}_g f^{bg}{}_a X^j_b X^i_c X^j_d + {\cal O}\left(\frac{1}{g_{YM}^2}\right)\,,
\nonumber\\
0 &=& \nabla^\mu F_{\mu\nu\phi a} - \frac{1}{2} \left(  X_c^i \nabla_\nu X_d^i + \frac{i}{2} \bar\Psi_c \Gamma_\nu \Psi_d\right)f^{cd}{}_a + {\cal O}\left(\frac{1}{g_{YM}^2}\right)\,, \nonumber
\end{eqnarray}
where $X^I_{\phi}=(X^i_\phi,x^8_\phi)$ with $i=1,... ,7$, the covariant derivative $\nabla_\mu$ is defined as
\begin{equation}
(\nabla_\mu X)_a = \partial_\mu X_a - \tilde A_\mu{}^b{}_a X_b = \partial_\mu X_a - 2A_{\mu\phi b} X_c f^{bc}{}_a
\end{equation}
and $F_{\mu\nu}{}^B{}_A$ is defined in (\ref{Fwot}). Note that the Higgs mechanism has transformed the algebraic equation for $\tilde A_{\mu\nu}{}^\phi{}_a$ into a dynamical equation for the gauge potential $A_{\mu\phi a}$ \cite{Mukhi:2008ux}.

The leading terms in (\ref{D2eqs})  can be integrated to the Lagrangian
\begin{equation}
{\cal L} = \frac{1}{g_{YM}^2} \Big({\cal L}_{\rm decoupled} + {\cal L}_{\rm 0}\Big)\,,
\end{equation}
where
\begin{equation}
{\cal L}_{\rm decoupled} = -\frac{1}{2}\partial_\mu X_\phi^I \partial^\mu X_\phi^I + \frac{i}{2} \bar\Psi_\phi \Gamma^\mu \partial_\mu \Psi_\phi
\end{equation}
and ${\cal L}_0$ is the standard $2+1$ dimensional SYM Lagrangian
\begin{eqnarray}
{\cal L}_0 &=& -\frac{1}{4} F_{\mu\nu a}F^{\mu\nu a} -\frac{1}{2}\nabla_\mu X^{a i} \nabla^\mu X_a^i + \frac{1}{4} \left( f_{abc} X^{a i} X^{b j}\right) \left( f_{de}{}^c X^{d i} X^{e j}\right)
\nonumber\\
&& + \frac{i}{2}\bar \Psi^a \nablaslash \Psi_a + \frac{i}{2} f_{abc} \bar \Psi^a \Gamma^i X^{b i} \Psi^c \,,
\end{eqnarray}
where we have here denoted $F_{\mu\nu a}=4 F_{\mu\nu \phi a}$.

In accordance with \cite{Mukhi:2008ux}, the scalars $X^{8}_{a}$ behave as Goldstone bosons giving a mass to $\tilde A_{\mu\nu}{}^\phi{}_a$. Moreover, the scalar  $x^8_\phi$ in (\ref{fluct}), corresponding to the fluctuation around the vev, can be dualised to an abelian gauge field and will, together with the centre of mass modes $X^i_{\phi}$ for the D2-branes and the superpartners $\Psi_{\phi}$, form a free $U(1)$ vector multiplet. In addition to the fact that this $U(1)$ centre of mass multiplet is decoupled to lowest order in $g_{YM}^{-1}$, in our construction it actually obeys free dynamics to all orders in $g_{YM}^{-1}$.

\section{Conclusions}\label{concl}

In light of the fact that it seems to be exceedingly difficult to find solutions to the fundamental identity based on a totally antisymmetric $f^{ABCD}$, which is required for a Lagrangian description, we work instead at the level of field equations. Since there is no \emph{a priori} reason to expect a Lagrangian formulation for the theories which describe conformal fixed points of the nonconformal $\mathcal{N}=8$ SYM theories it seems relevant to not rule out this possibility. 

Putting potential problems that could accompany this approach aside for the moment, we use a choice of $f^{ABC}{}_D$ that satisfies the fundamental identity but is not totally antisymmetric. With this choice we are able to obtain, using the Higgs mechanism of \cite{Mukhi:2008ux}, the SYM field equations for any (compact and semi-simple) gauge group, to leading order in $g_{YM}^{-1}$. A novel feature of this construction is that the $U(1)$ centre of mass multiplet for the D2-branes obeys free dynamics, even before applying the Higgs mechanism and breaking the $SO(8)$ covariance. This means that we have identified a candidate for the centre of mass multiplet also for multiple M2-branes.
We believe that these results indicate that this approach could be consistent despite the lack of a Lagrangian, but this of course needs further investigation. Note that, the leading order terms, corresponding to the ordinary SYM D2-theory, of course do have a Lagrangian formulation.

An interesting feature of this construction is that the choice of $f^{ABC}{}_D$ in (\ref{ansats}) seems rather unique if we want to make contact with D2-brane physics. In order to have a $U(1)$ centre of mass multiplet obeying free dynamics we need to set $f^{ABC}{}_\phi=0$. Then, out of the remaining components we need to embed the Lie algebra structure constants, implying that $f^{\phi a b}{}_c \sim f^{ab}{}_c$. The only remaining components, $f^{abc}{}_d$ , are forced to vanish by the fundamental identity. Thus, since the choice of $f^{ABC}{}_D$ seems to be so tightly constrained, it would be interesting to also investigate what this proposed D2-M2 correspondence implies to higher order in $g_{YM}^{-1}$.

\paragraph{Note added:}
As this paper was being prepared, a preprint appeared \cite{Morozov:2008cb} which has some overlap with our paper.

\acknowledgments

We would like to thank Martin Cederwall, Gabriele Ferretti, Andreas Gustavsson, Sunil Mukhi and Per Salomonsson for discussions. The work of U.G.~is funded by the Swedish Research Council.

\appendix

\section{Useful formul\ae}\label{useful}

In this appendix we collect some useful formul\ae\ needed for the computations related to the Higgs mechanism. We restrict our attention to compact and semi-simple Lie algebras, which means that the generators can be chosen to satisfy
\begin{equation}
{\rm tr} (T^a T^b) = \lambda \, \delta^{ab} \,, \label{lambda}
\end{equation}
where
\begin{eqnarray}
(T^a)^b{}_c &=& - i f^{ab}{}_c \,,
\nonumber\\
\,[ T^a,T^b ] &=& i f^{ab}{}_c T^c
\end{eqnarray}
and the structure constants $f^{ab}{}_c$ are totally antisymmetric. In these conventions the Jacobi identity is
\begin{equation}
f^{[ab}{}_d f^{c]d}{}_e =0 \,.
\end{equation}

From the constraint $\tilde A_\mu{}^B{}_A = A_{\mu CD}f^{CDB}{}_A$, and the above Lie algebra properties, we find that
\begin{equation}
\tilde F_{\mu\nu}{}^d{}_c f_{e}{}^c{}_d f_{e}{}^b{}_a  = - \lambda \, \tilde F_{\mu\nu}{}^b{}_a \,,
\end{equation}
which is needed for the Higgsing. Note that in the final formul\ae\, (\ref{D2eqs}), the normalisation constant $\lambda$ drops out.


\begin{thebibliography}{999}

\bibitem{Schwarz:2004yj}
  J.~H.~Schwarz,
  ``Superconformal Chern-Simons theories,''
  JHEP {\bf 0411} (2004) 078
  [arXiv:hep-th/0411077].
  %%CITATION = JHEPA,0411,078;%%

\bibitem{Basu:2004ed}
  A.~Basu and J.~A.~Harvey,
  ``The M2-M5 brane system and a generalized Nahm's equation,''
  Nucl.\ Phys.\  B {\bf 713} (2005) 136
  [arXiv:hep-th/0412310].
  %%CITATION = NUPHA,B713,136;%%

\bibitem{Bagger:2006sk}
  J.~Bagger and N.~Lambert,
  ``Modeling multiple M2's,''
  Phys.\ Rev.\  D {\bf 75} (2007) 045020
  [arXiv:hep-th/0611108].
  %%CITATION = PHRVA,D75,045020;%%

\bibitem{Gustavsson:2007vu}
  A.~Gustavsson,
  ``Algebraic structures on parallel M2-branes,''
  arXiv:0709.1260 [hep-th].
  %%CITATION = ARXIV:0709.1260;%%

\bibitem{Bagger:2007jr}
  J.~Bagger and N.~Lambert,
  ``Gauge Symmetry and Supersymmetry of Multiple M2-Branes,''
  Phys.\ Rev.\  D {\bf 77} (2008) 065008
  [arXiv:0711.0955 [hep-th]].
  %%CITATION = PHRVA,D77,065008;%%

\bibitem{Bagger:2007vi}
  J.~Bagger and N.~Lambert,
  ``Comments On Multiple M2-branes,''
  JHEP {\bf 0802} (2008) 105
  [arXiv:0712.3738 [hep-th]].
  %%CITATION = JHEPA,0802,105;%%

\bibitem{Gustavsson:2008dy}
  A.~Gustavsson,
  ``Selfdual strings and loop space Nahm equations,''
  arXiv:0802.3456 [hep-th].
  %%CITATION = ARXIV:0802.3456;%%

\bibitem{Mukhi:2008ux}
  S.~Mukhi and C.~Papageorgakis,
  ``M2 to D2,''
  arXiv:0803.3218 [hep-th].
  %%CITATION = ARXIV:0803.3218;%%

\bibitem{Bandres:2008vf}
  M.~A.~Bandres, A.~E.~Lipstein and J.~H.~Schwarz,
  ``N = 8 Superconformal Chern--Simons Theories,''
  arXiv:0803.3242 [hep-th].
  %%CITATION = ARXIV:0803.3242;%%

\bibitem{Berman:2008be}
  D.~S.~Berman, L.~C.~Tadrowski and D.~C.~Thompson,
  ``Aspects of Multiple Membranes,''
  arXiv:0803.3611 [hep-th].
  %%CITATION = ARXIV:0803.3611;%%

\bibitem{VanRaamsdonk:2008ft}
  M.~Van Raamsdonk,
  ``Comments on the Bagger-Lambert theory and multiple M2-branes,''
  arXiv:0803.3803 [hep-th].
  %%CITATION = ARXIV:0803.3803;%%

\bibitem{Lambert:2008et}
  N.~Lambert and D.~Tong,
  ``Membranes on an Orbifold,''
  arXiv:0804.1114 [hep-th].
  %%CITATION = ARXIV:0804.1114;%%

\bibitem{Distler:2008mk}
  J.~Distler, S.~Mukhi, C.~Papageorgakis and M.~Van Raamsdonk,
  ``M2-branes on M-folds,''
  arXiv:0804.1256 [hep-th].
  %%CITATION = ARXIV:0804.1256;%%

\bibitem{FigueroaO'Farrill:2002xg}
  J.~Figueroa-O'Farrill and G.~Papadopoulos,
  ``Pluecker-type relations for orthogonal planes,''
  arXiv:math/0211170.
  %%CITATION = MATH/0211170;%%
  
\bibitem{Bergshoeff:2003ri}
  E.~Bergshoeff, U.~Gran, R.~Linares, M.~Nielsen, T.~Ortin and D.~Roest,
  ``The Bianchi classification of maximal D = 8 gauged supergravities,''
  Class.\ Quant.\ Grav.\  {\bf 20} (2003) 3997
  [arXiv:hep-th/0306179].
  %%CITATION = CQGRD,20,3997;%%

\bibitem{Awata:1999dz}
  H.~Awata, M.~Li, D.~Minic and T.~Yoneya,
  ``On the quantization of Nambu brackets,''
  JHEP {\bf 0102} (2001) 013
  [arXiv:hep-th/9906248].
  %%CITATION = JHEPA,0102,013;%%
  
\bibitem{Morozov:2008cb}
  A.~Morozov,
  ``On the Problem of Multiple M2 Branes,''
  arXiv:0804.0913 [hep-th].
  %%CITATION = ARXIV:0804.0913;%%




\end{thebibliography}
\end{document}